\documentclass[11pt,letter]{article}
\usepackage{jheppub}
\allowdisplaybreaks
\usepackage{color}
\usepackage{amsmath}
\usepackage{graphicx}
\usepackage{esint}
\usepackage{slashed}
\usepackage[english]{babel}
\usepackage{hyperref}
\usepackage[normalem]{ulem}
\usepackage{comment}
\usepackage[utf8]{inputenc}

\usepackage{babel}
\begin{document}

\title{ Neutrino Mass from  Affleck-Dine  Leptogenesis and WIMP Dark Matter }

\author{Rabindra N. Mohapatra$^a$ and}
\author{Nobuchika Okada$^b$}
\affiliation{ $^a$Maryland Center for Fundamental Physics and Department of Physics, University of Maryland, College Park, Maryland 20742, USA}
\affiliation{ $^b$Department of Physics, University of Alabama, Tuscaloosa, Alabama 35487, USA}
\emailAdd{rmohapat@umd.edu}
\emailAdd{oakadan@ua.edu}

\date{\today}

\abstract
{Affleck-Dine (AD)  mechanism for leptogenesis involves the cosmological evolution of a complex scalar field (AD field) 
that carries non-zero lepton number. 
We show how explicit lepton number breaking terms, which involve the AD field needed to implement this scenario combined 
with fermionic WIMP dark matter, can generate  neutrino mass at the one loop level, thus providing a unified framework 
for solving four major puzzles of the standard model i.e. inflation, baryogenesis, dark matter and neutrino mass.   
We discuss some phenomenological implications of this model.}


\maketitle

\section{Introduction} 
Understanding the origin of neutrino masses and the matter-antimatter asymmetry are two of the major challenges facing particle theory research today. The solutions to these puzzles  together with  the unravelling of the mystery of the dark matter, will be crucial windows to physics  beyond the standard model (BSM). While there are many mechanisms proposed in the literature for solving  these problems separately, unified approaches to them within a single theoretical framework, in addition to being more appealing, are expected to  provide deeper insight into the BSM landscape and are therefore worth pursuing.  An additional advantage of such unified frameworks is  that unification can lead to testable constraints on the parameters of the model.

One well known example of  a partially unified scenario is the proposal of leptogenesis~\cite{FY}
 which is based on the seesaw mechanism for understanding neutrino masses~\cite{seesaw1,seesaw2,seesaw3,seesaw4,seesaw5}. 
 In this kind of an approach, the mechanism for understanding neutrino masses  leads to an understanding of matter-antimatter asymmetry. 
 However, dark matter remains outside of typical frameworks of this type and require separate physics.
  
  In this paper, we take a different approach and start with the Affleck-Dine mechanism~\cite{AD} to create the lepton asymmetry  and show how this provides a reverse path where AD leptogenesis and a WIMP dartk matter leads to neutrino masses at the one loop level. The basic idea goes as follows:
  Typically, in the AD mechanism, one relies on a lepton number carrying complex scalar field (called here AD field and denoted here by $\Phi$) with the Lagrangian of the model explicitly breaking lepton number ($L$) by a quadratic term in the $\Phi$ field. 
  In the presence of this $L$ breaking term, the cosmological evolution of $\Phi$ generates lepton asymmetry.  We point out that the reverse path for understanding neutrino masses in this case comes from the same lepton number breaking $\Phi^2$ term in the Lagrangian, in combination with a fermionic WIMP dark matter  as we show below.
 Thus,  neutrinos masses are a consequence of AD leptogenesis plus dark matter. Of course, neutrinos in this kind of scenario are naturally Majorana type fermions.  

We further note that while the inflaton and the AD fields are separate fields~\cite{AD,DRT,mazu1,mazu2} in many AD scenarios, there are examples where the inflaton field and the AD field can be identified thus providing unification of inflation and baryo/leptogenesis~\cite{Cline:2019fxx, Charng:2008ke, Hertzberg:2013jba, Takeda:2014eoa, Lin:2020lmr, stubbs, russian, Kawasaki:2020xyf, Barrie:2021mwi, nobu}.  
We adopt one such scenario here~\cite{stubbs, russian} so that we indeed have a unified framework for four of the puzzles of the standard model: inflation, baryogenesis, dark matter and neutrino masses.

 In our recent work~\cite{nobu2}, another unified scenario was presented by using a similar framework, where spontaneous  breaking of the global charge, the  $L$  symmetry carried by the inflaton field was used to generate neutrino masses. There, the neutrino masses arose at the tree level using the inverse seesaw mechanism and the dark matter was a consequence of this spontaneous breaking of lepton number, giving rise to a pseudo-Goldstone dark matter. This required that the inflaton and AD field acquire a vacuum expectation value  (vev)  putting additional constraints on the model. 
 In the new scenario discussed in the present paper,  the inflaton field does not have a vev and  neutrino mass arises as a radiative effect
 (for a comprehensive review on models radiatively generating neutrino masses, see  Ref.~\cite{volkas}) 
from the  already present term that breaks $L$ symmetry explicitly. 
The model has an automatic  $Z_2$ symmetry that guarantees the stability of dark matter.
 We also discuss implications of  a possible supersymmetric embedding of this model. 

This paper is organized as follows: in sec.~2, we present an outline of the model and isolate its symmetries; 
in sec.~3, we discuss the evolution of the universe in this picture, and discuss leptogenesis in sec~4. 
In sec.~5 we focus on the one loop generation of neutrino mass;
in sec.~6, we discuss the constraints on the model parameters and provide two benchmark set and in sec.~7, 
dark matter candidate in the model is discussed; 
in sec. 8, we comment on possible phenomenological implications of this model and
sec.~9 is devoted to a summary of the results.

\section{ The model} 

The model is based on the standard model (SM) gauge group $SU(3)_c \times SU(2)_L\times U(1)_Y$.
The particle content is listed in Table~\ref{tab1} 
In addition to the  SM particle content, we introduce the new fields i.e.~an AD field $\Phi$, which is SM singlet and carries a lepton number $+1$, three fermionic doublets $D_i$ ($i=1, 2, 3$) with hypercharge $Y=-1$ and zero lepton number and their mass partners $\bar{D}_i$ 
with $Y=+1$ and $L=0$, three $L=0$ fermionic singlets $\chi_i$. 
The presence of the $D$ and $\bar{D}$  together makes the model anomaly free.

\begin{table}[t]
\begin{center}
\begin{tabular}{|c||c||c||c|}
\hline
Field & $U(1)_L$& SM quantum number & $Z_2'$ \\ \hline
Fermion &  &  &\\
$\ell_a$ & $+1$& $({\bf 2},-1)$ & + \\
$e^c_a$ & $-1$ & $({\bf 1}, +2)$& +\\
$D_i$ & $0$& $({\bf 2},-1)$&$-$\\
$\bar{D}_{i}$& $0$ & $({\bf 2},+1)$& $-$\\
$\chi_i$&$0$ & $({\bf 1},0)$&$-$\\\hline
Scalar & & & \\
$H$ &$0$ & $({\bf 2},+1)$&+ \\
$\Phi$ & $-1$& $({\bf 1},0)$&$-$ \\\hline
\hline
\end{tabular}
\caption{
Particle content  of the model responsible for one loop neutrino mass and dark matter.  
$D_i$, $\bar{D}_i$ and $\chi_i$ are new fermionic fields as stated in the text. 
$H$ is the SM Higgs doublet. 
The subscript $a$  goes over lepton flavors and $i$ goes over $D$ flavors with $a, i=1,2,3$.
 The  lepton number of the different fields under $U(1)_L$  are shown in the second column. 
 The SM $SU(2)_L\times U(1)_Y$ quantum numbers are in the third column.  
 The $Z^\prime_2$ quantum numbers in the table are guaranteed by the $U(1)_L$ symmetry. 
 The $SU(3)_c$  group has been suppressed and all fields shown are color singlets. 
 We have also omitted the quark fields.
}
\end{center}
\label{tab1}
\end{table}

The Lagrangian of the model is given by
\begin{eqnarray}
{\cal L} &=& {\cal L}_{SM}+{\cal L}_{inf} (\Phi, R)
+ (Y_{\Phi})_{ai} \ell_a\bar{D}_i \Phi+ (Y_{D})_{ij} D_i \chi_j H \nonumber \\
&&+(Y_{\bar{D}})_{ij} \bar{D}_i \chi_j \tilde{H} + \mu_{ij}\chi_i\chi_j+ (m_D)_{ij} D_i\bar{D}_j +h.c. \nonumber\\
&&
+(\partial_\mu \Phi)^\dagger (\partial^\mu \Phi)
 -\left( m^2_\Phi |\Phi|^2+\lambda |\Phi|^4+\epsilon m^2_\Phi (\Phi^2+\Phi^{\dagger 2} ) \right) ,
\label{eq:L}
\end{eqnarray}
where $\tilde{H}=i\tau_2 H^*$, and $m^2_\Phi > 0$;  
$ {\cal L}_{SM}$ is the SM Lagrangian, 
${\cal L}_{inf}$ denotes the non-minimal $\Phi$ coupling to gravity of the form 
${\cal L}_{inf}= -\frac{1}{2} (M^2_P + \xi |\Phi|^2) R$ (see, for example, Refs.~\cite{inf1,inf2}) 
that plays a crucial role for the successful inflation, 
where $R$ is the Ricci scalar, and $M_P=2.4 \times 10^{18}$ GeV is the reduced Planck mass.  
As shown in Table I, the Lagrangian has the global symmetry $U(1)_L$ explicitly broken by the $\epsilon$ term.  The model also has an additional  $Z'_2$ symmetry under which the fields $\Phi, \chi, D, \bar{D}$ are odd and the rest of the fields are even.This $Z'_2$ symmetry allows for the existence of a  fermionic dark matter, which is a linear combination of the neutral components of the lightest of the $D_i$ fields $D_1^0, \bar{D}^0_1$ and $\chi$ fields. We discuss this in a subsequent section.
 
 We will also see in a subsequent section,  that this Lagrangian leads to
a one loop Majorana mass for neutrinos proportional to $\epsilon$ whereas the baryon to entropy ratio generated by the AD mechanism 
gives $n_B/s$ is inversely proportional to $\epsilon$ thereby relating the neutrino mass with the lepton asymmetry in a way different from leptogenesis.



\section{  Inflation and evolution of the AD field } 
To implement AD leptogenesis in the model,  we need to study the evolution of the AD field 
till the epoch when it $H\simeq m_\Phi$. This has been discussed earlier in~\cite{stubbs, nobu}.


First  stage in the evolution of the inflaton/AD field $\Phi$ is when $\Phi$ field has a value larger than $M_P/\sqrt{\xi}$, 
so that its non-minimal coupling to gravity causes inflation (see Ref.~\cite{nobu} for this discussion
where the earlier work has been reviewed). 
The non-minimal coupling of the $\Phi$ field to gravity helps to implement inflation
in identifying the AD field with inflaton.

The inflation is characterized by a parameter $\xi$ which denotes coupling of $\Phi$ to the Ricci scalar.
This model is known to fit all the Planck 2018 data on the spectral index and the tensor-to-scalar ratio. 
The $\Phi$ then slowly rolls down the potential and inflation comes to an end as $\Phi$ becomes less than $M_P/\sqrt{\xi}$. The $\Phi $ field subsequently decreases like $1/a(t)$, where $a(t)$ is the scale factor of the universe, until its value becomes below $m_\Phi/\sqrt{\lambda_\Phi}$. 
At this point, the oscillation of the $\Phi$ field starts separately for its real and imaginary parts
defined as $\Phi=\frac{1}{\sqrt{2}} (\phi_1 +i \phi_2)$, whose initial values are different. 
 They evolve starting from two random values for the two parts. 
This difference between the initial values of $\phi_1$ and $\phi_2$, introduces the CP violation required by the Sakharov's criterion
  for baryo/leptogenesis. 
The oscillation of the AD field leads to an asymmetry in the abundance of $\ell D$ and $\bar{\ell} \bar{D}$ 
  which is generated when the AD field decays as $\Phi \to \ell D$.  This decay reheats the universe to  temperature $T_R\simeq \sqrt{\Gamma_\Phi M_P}$, which must be less than the $\Phi$ mass for the generated lepton asymmetry to survive. This leads to a constraint on the model parameters which we quantify later. For now, we define $T_R=Km_\Phi$ with a constant  $K$. 

We estimate the reheat temperature using the formula $T_R\simeq \sqrt{\Gamma_\Phi M_P}$, 
  where $\Gamma_\Phi$ is the total decay width of the inflaton/AD field. 
To calculate the total $\Phi$ decay width, we assume the following mass hierarchy among the $D_{1,2,3}$ i.e. $m_{D_1}\ \ll m_\Phi < m_{D_{2,3}}$. 
  With this choice of mass arrangement, i.e.~${D_{2,3}}$  do not contribute to the  decay of $\Phi$ and the total $\Phi$ decay width is given by
  \begin{eqnarray}
  \Gamma_\Phi =\sum_a \Gamma_{\ell_a D_1}\simeq \frac{Y^*_{a1}Y_{a1}}{4\pi} m_{\Phi}, 
  \end{eqnarray}
 where $a$ goes over all lepton flavors.  As we will see below, in our model of one loop neutrino masses, $\sum_a Y^*_{a1}Y_{a1}\propto m_{1}$, 
 where $m_1$ is the mass of the lightest neutrino independent of the  flavor structure in the $Y_D$
  and therefore without constraining any other neutrino oscillation observable except $m_{1}$ (which is unknown), 
  we can get a $T_R =  Km_\Phi$ with $K< 1$.  
In the process, we will find an upper limit for $ m_1$ which can provide a test of the model, once neutrinoless double beta decay is discovered.



\section{ Lepton asymmetry generation}
Coming to generation of lepton asymmetry,  we note that the difference between the initial values of $\phi_1$ and $\phi_2$ introduces the CP violation
  required by the Sakharov's criterion for baryo/leptogenesis and leads to lepton asymmetry
  when $\Phi$ decays to the  $\ell+D$ via $\Phi\to \ell_a D_1$ process when the inflaton field starts oscillating  and reheats the universe to the temperature $T_R$ noted above. 
We choose parameters such that $T_R =K m_\Phi$ with $K< 1$, as discussed above.  
In the next section we will see the constraints imposed by this requirement on our model. 

We  first note that in such a leptogenesis scenario, the lepton number to entropy ratio is given by~\cite{stubbs} 
\begin{eqnarray}
\frac{n_L}{s}~\simeq \frac{T_R^3}{\epsilon \, m^2_\Phi \,  M_P} \simeq 10^{-10}.
\label{nB}
\end{eqnarray}
The conditions under which this equation holds are that $\epsilon \ll 1$ and $\epsilon m_\Phi/\Gamma_\Phi\gg 1$. Both these conditions are satisfied in our model. 

An important input into this estimate of $n_B/s$ is the reheat temperature $T_R=KM_P$, 
which must be less than the AD field mass $m_\Phi$, i.e.~$K< 1$ as already noted.
This implies the following relation between $m_\Phi$, $\epsilon $ and $K$ i.e.
\begin{eqnarray}
m_\Phi \simeq 10^{-10}\frac{\epsilon}{K^3} M_P. 
\label{nBsK}
\end{eqnarray}
\begin{figure}[tb]
  \centering
 \includegraphics[width=0.7\linewidth]{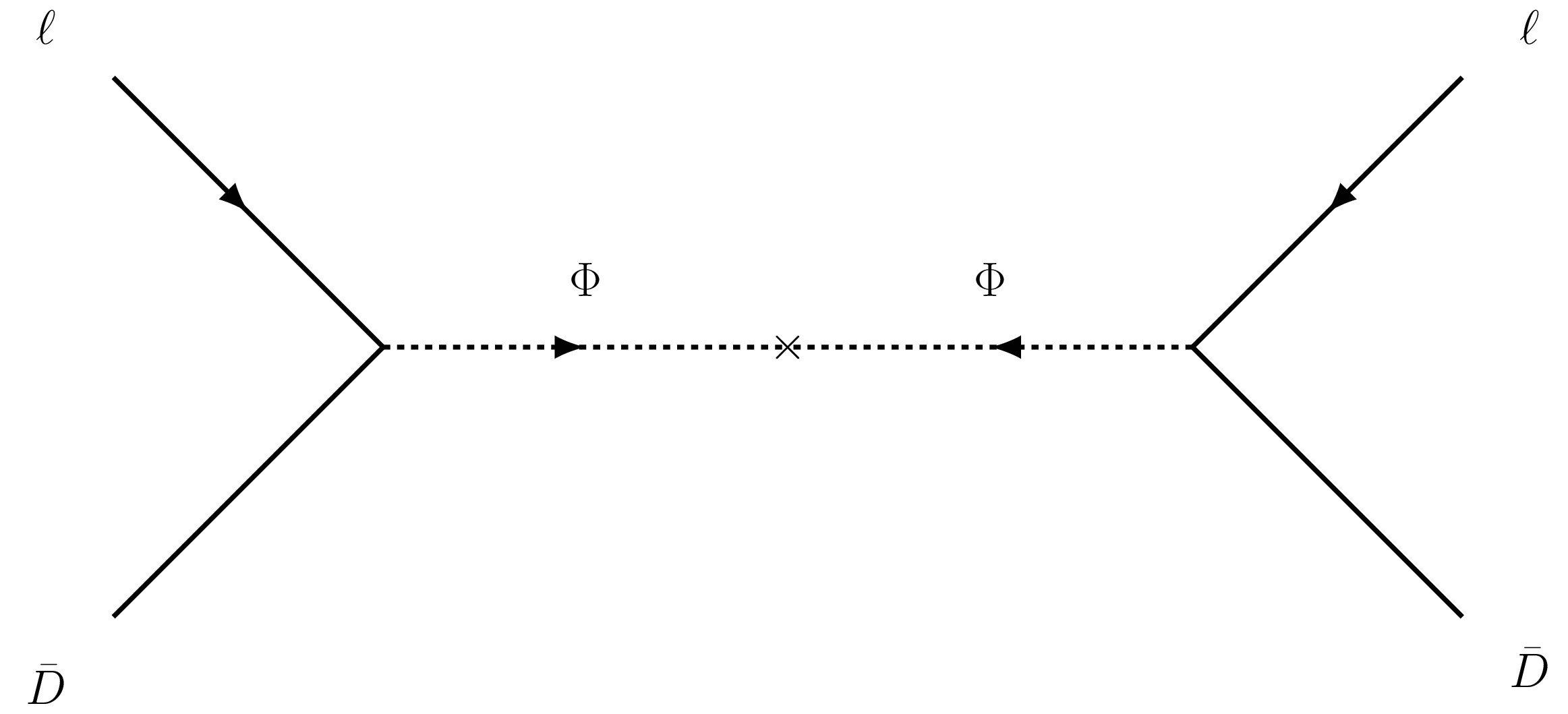}
  \caption{Feynman diagram responsible for washout of lepton asymmetry.
  The arrows indicate the flow of the lepton number.} 
  \label{fig1}
  \end{figure}

The model has explicit lepton number violating interaction given by $\epsilon$ and  can cause the lepton asymmetry generated to be washed out unless its value is small enough. We determine this value below.
The lepton asymmetry washout in our model can be caused by the lepton number breaking term in Eq.~(\ref{eq:L}) proportional to $\epsilon$.
Let us now look at the constraints imposed on the model by the fact that some interactions such as  $\epsilon m^2_\Phi$ term 
in the potential breaks $L=2$ and can cause  washout unless it is out of equilibrium above $T_R$.
To establish the constraints on the model due to this, we look at the scattering $\ell D\leftrightarrow  \ell^* D^*$ 
mediated by the $\Phi$ exchange and the $\epsilon$ term (see Fig.~\ref{fig1}) since it breaks $L$ by two units.
We demand that this process be out
of equilibrium above $T_R$ and find the condition,

\begin{eqnarray}
  T_R^3 \times \frac{Y_\Phi^4}{4 \pi} \frac{\epsilon^2 T_R^2}{m_\Phi^4} < H = \sqrt{\frac{\pi^2}{90} g_{*} } \frac{T_R^2}{M_P},
  \label{wash}
\end{eqnarray}
where $g_* \simeq 100$ is the effective degrees of freedom of the SM thermal plasma.  We discuss the implication of this constraint for parameters of the model in Sec.~6.

\section{ One loop neutrino mass} 
\begin{figure}[tb]
  \centering
 \includegraphics[width=0.7\linewidth]{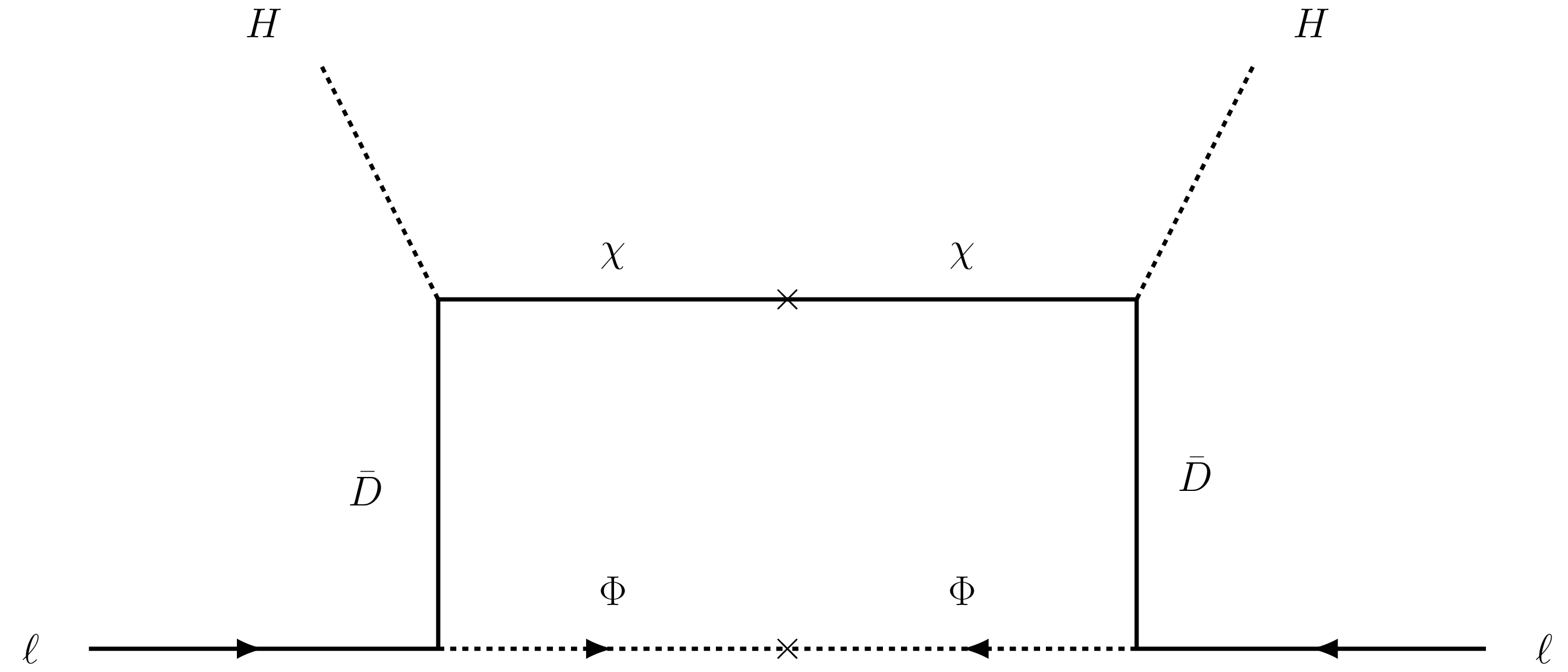}
  \caption{
Feynman diagram responsible for one loop neutrino mass.
Arrows indicate the flow of the lepton number. 
The upper cross denotes the Majorana mass insertion of $(\mu)_{ij}$
while the lower cross is for the insertion of $\epsilon m_\Phi$.
} 
  \label{fig2}
  \end{figure}

We now discuss how neutrino masses and mixings can arise in this model and the consistency with observations. 
There is no tree level neutrino mass in this model. It arises at the one loop level from the diagram shown in Fig.~\ref{fig2}.\footnote{
There is another one loop diagram with the Yukawa coupling $(Y_{\bar D})_{ij}$ in Eq.~(\ref{eq:L}). 
For simplicity, we assume $(Y_{\bar D})_{ij}$ is negligibly small. 
} 
To discuss this contribution, we choose a basis without loss of generality. 
In this basis, $Y_D$ and $\mu$ are diagonal and $Y_\Phi$ is a full matrix with all non-zero elements. 
By a suitable choice of basis we can also make $m_{D}$ diagonal. 
In this case, the one loop induced neutrino mass can be written as (see Fig.~\ref{fig2})
\begin{eqnarray}
(m_\nu)_{ab} \simeq \frac{\left(Y_\Phi Y_D \mu Y^T_D Y^T_\Phi \right)_{ab}}{16\pi^2}\frac{v^2_{wk}}{m^2_\Phi},  
\end{eqnarray}
where $v_{wk}$ is the SM Higgs vev, and we have assumed $m_{D_{2,3}}$ is the same order of $m_\Phi$. 
For simplicity, we make a further assumption that $Y_D$ and $\mu$ are flavor universal
together with choice $\mu \simeq m_\Phi {\bf I}$ ($\bf {I}$  being the unit matrix) and $(Y_D)_{ij}= Y_D\delta_{ij}$.
We can  then write the neutrino mass matrix  as
\begin{eqnarray}
(m_{\nu})_{ab}\simeq \frac{\epsilon \,  v^2_{wk}}{16\pi^2 m_\Phi} \, (Y_{\Phi} Y^T_{\Phi} )_{ab}  \, Y^2_D.
\end{eqnarray}
Using Eq.~(\ref{nBsK}), we can write the above expression for the neutrino mass matrix $m_\nu$ as
\begin{eqnarray}
(m_\nu)_{ab}  \simeq \frac{10^{10}K^3  v^2_{wk}}{16 \pi^2 M_P} \, (Y_{\Phi} Y^T_{\Phi} )_{ab} \, Y^2_D.
\label{nu1}
\end{eqnarray}
From the neutrino oscillation data, we set the light neutrino mass eigenvalue to be  $m_{2,3}\sim 10^{-10}$ GeV
for the normal hierarchy, so that Eq.~(\ref{nu1}) has the implications that  the Yukawa couplings satisfy the following condition:
 \begin{eqnarray}
 (Y_\Phi Y^T_\Phi)_{ab} (Y_D)^2 \simeq \frac{10^{-4}}{K^3}. 
 \label{yphi}
 \end{eqnarray}
where $Y_{\Phi}$ couplings in Eq.~(\ref{yphi}),  refer to the Yukawa couplings of the second and third generation leptons. The  requirement from perturbativity i.e.~all $Y_{\Phi,D}\lesssim 1$ implies that  $K^3 \gtrsim 10^{-4}$.

 Let us now discuss the total $\Phi$ decay width which is clearly related to the neutrino mass matrix. 
 Note that we have from  $m_\nu=U^* D_\nu U^\dagger$,  
 \begin{eqnarray}
 (Y_\Phi)_{ai} =\frac{1}{\sqrt{X}}( U^*\sqrt{D_\nu})_{ai}, 
 \end{eqnarray}
 where $X= \frac{\epsilon v^2_{wk}}{16\pi^2 m_\Phi} Y_D^2$,
 and $D_\nu ={\rm diag}(m_1, m_2, m_3)$.
From this equation, we find 
 \begin{eqnarray}
 \Gamma_\Phi=\sum_a\Gamma_{\Phi\to \ell_a D_1}=(Y^\dagger_{\Phi} Y_{\Phi})_{11}\frac{m_\Phi}{4\pi} 
= \frac{m_\Phi}{4 \pi X} m_1. 
 \end{eqnarray}
We then use $T_R \simeq K m_\Phi =\sqrt{\Gamma_\Phi M_P}$ to get 
 \begin{eqnarray}
 m_1({\rm eV}) \simeq 10^{-6}\times Y^2_D \, \epsilon \, K^2. 
 \end{eqnarray}
 We thus see that for $Y_D\sim 1$,  the lightest neutrino mass has to be 
 $m_1  \ll 10^{-6}$ eV for $\epsilon \ll 1$ and $K<1$. 

\section{ Collection of constraints and two benchmark sets of parameters}
In this section, we collect the constraints on the various parameters of the model that follow from neutrino mass generation, adequate leptogenesis and acceptable reheat temperature $T_R$ . The constraints are:
\begin{eqnarray}
m_\Phi \simeq 10^{-10}\frac{\epsilon}{K^3} M_P
\end{eqnarray}
and the one loop neutrino masses that are expected from oscillation data for the case of normal hierarchy  and perturbativity of Yukawa couplings is given by
\begin{eqnarray}
K^3\gtrsim 10^{-4}.
\end{eqnarray}
 Once this condition is satisfied, any choice of $K, \epsilon\ll 1$ and $m_\Phi$ works to yield the right $n_B/s$ and required $m_\nu$ values.

For our parameterization, the no washout condition in Eq.~(\ref{wash})  translates to
\begin{eqnarray}
m_\Phi \gtrsim K^3 \frac{Y^4_\Phi \epsilon^2}{4\pi} M_P \simeq \frac{10^{-8} \epsilon^2}{4\pi K^3}M_P 
\end{eqnarray}
by using Eq.~(\ref{yphi}) with $Y_D \sim 1$. 
This lower limit for $\Phi$ mass is in terms of parameters $\epsilon$ and $K$; so combining it with Eq.~(\ref{nBsK}), we get
\begin{eqnarray}
\epsilon \lesssim 4\pi \times 10^{-2}, 
\end{eqnarray}
which is consistent with our assumption of $\epsilon \ll 1$ in the model. 
In Table~\ref{tab2}, we give two sets of benchmark points (and clearly, the points in between) 
that satisfies the requirements of the model. We see that the model has an ample parameter space 
where all the physical requirements can be satisfied.

\begin{table}[t]
\begin{center}
\begin{tabular}{|c||c||c|}\hline
parameter &value(set 1) &value(set 2)\\\hline
$\epsilon$ & $10^{-5}$ & $10^{-3}$ \\
$K$ & $0.1$ & $0.1$ \\
$m_\Phi$& $\sim 10^6$ GeV& $\sim 10^8$ GeV\\
$m_{D_1}$ & $10^3 ~{\rm GeV}$  & $10^3~ {\rm GeV}$ \\
$m_{D_{2,3}}$ & $\sim 3\times 10^6$ {\rm GeV}&$\sim 3\times 10^8$ GeV\\
$(Y_{\Phi})_{a1}$ ($a=1,2,3$) & $\sim 10^{-6.5}$ & $\sim 10^{-5.5}$\\
$(Y_{\Phi})_{ai}$ ($a=1,2,3$; $i \neq 1$) & $\sim 10^{-0.5}$ &$\sim 10^{-0.5}$\\ \hline
\end{tabular}
\end{center}
\caption{Two sets of benchmark parameters that satisfy all the constraints considered in the model. They cover all points in between and thus represent a broad parameter space of the model.
}
\label{tab2}
\end{table}

\section{ Dark matter in the model} 
We will see in this section that the lightest of the fermionic $D$ and $\bar{D}$ particles in the model is stable and  can play the role of dark matter in the universe. Two things are worth 
noting to appreciate this point: first that there is a discrete symmetry in the Lagrangian (called $Z'_2$ above, which is analogous to R-parity) under which $\Phi$, $D_a$, $\bar{D}_a$ and $\chi_a$ are odd and the rest of the particles are even. Note however that the  $Z'_2$ odd states $D_{2,3}$ can decay to $D_1$, there is no lighter states with odd $Z_2$ parity to which the DM can decay. The DM can however annihilate to standard model particles to give the right relic density of the universe.

To discuss these issues, let us decouple the the heavier fields $D_{2,3}$ and then identify the dark matter field. Note that the three particles $D^0_1, \bar{D}^0_1$ and $\chi$ fields mix after symmetry breaking and their mass matrix is given by
\begin{eqnarray}
M=\left(\begin{array}{ccc} D^0_1 & \bar{D}^0_1 & \chi \end{array}\right) \left(\begin{array}{ccc} 0 & m_{D_1} & Y_D v_{wk} \\ m_{D_1} & 0 & Y_Dv_{wk}\\  Y_D v_{wk} &  Y_D v_{wk} & \mu\end{array}\right)
\left(\begin{array}{c} D^0_1 \\ \bar{D}^0_1 \\ \chi \end{array}\right)
\end{eqnarray}
 It is clear from this mass matrix that the eigenstates are Majorana fermions and in the situation under consideration, $m_D , v_{wk}\ll \mu$ so that we can write the lightest eigenstate as
 \begin{eqnarray}
 \psi_{DM}\simeq \frac{1}{\sqrt{2}}D^0_1 +  \frac{1}{\sqrt{2}}\bar{D}^0_1 +\delta \, \chi  ,
 \end{eqnarray}
where $\delta\sim v_{wk}/\mu$, is a very small number since $v_{wk} \ll \mu\sim m_\Phi$ in our choice of parameters.
The structure is similar to the Higgsino-like neutralino DM in MSSM in the Wino and Bino decoupling limit. 
Thus the properties of our DM is essentially the same as the MSSM Higgsino-like DM scenario.
One implication of this is that, the annihilation process that gives the relic density is of the form 
$\psi_{DM} \psi_{DM}\to W^+W^-, ZZ, hh$. 
We estimate the thermally averaged cross section times relative velocity for this DM annihilation process as 
 \begin{eqnarray}
 \langle \sigma v_{rel} \rangle \sim \frac{g^4}{4\pi}\frac{1}{m^2_{DM}},
 \end{eqnarray}
 where $g$ is the $SU(2)_L$ gauge coupling. 
 This cross section  must be roughly one pico-barn to give the correct relic density implying that the  $m_{DM}\sim 1$ TeV.
 
 We can now discuss the direct detection cross section. The DM can scatter off a nucleus via the exchange of a $Z$ boson or Higgs boson of SM. Since the DM is a Majorana fermion, the $Z$-exchange contribution is spin dependent and the bounds on this are very weak. On the other hand, the Higgs boson exchange cross section is spin independent and can be large. The Feynman diagram for the Higgs exchange contribution to direct detection involves the $D$-component in the initial (or final) state and the $\chi$ component in the DM in the final (or initial) state, leading to the suppression factor $\delta$ in the amplitude. This cross section is therefore suppressed  since it is proportional to  $\delta^2$ in the parameter range of interest to us i.e. $m_{D_1}, v_{wk}\ll \mu$. This parameter region is called the ``blind spot" region 
 where the one loop graph is more important~\cite{blind}.

\section{ Comments} 
In this section, we make several comments on the model:

\begin{enumerate}

\item The one loop correction to the dynamics of the scalar field $\Phi$ comes from the couplings $\Phi\ell D$ or $Y_\Phi$ in Eq.~(\ref{eq:L}) and is of order $Y_\Phi^4/16 \pi^2$. We can choose $Y_\Phi \sim 1/3,$, which is quite compatible with Eq.~(\ref{yphi}) of the paper. 
In this case the one loop induced  $\Phi^4$ coupling is of order $10^{-4}$. For CMB fits, we may fix the tree-level $\lambda\Phi^4$ coupling 
to be of order $10^{-3}$. Therefore the one loop corrections are small and do not affect the scalar field dynamics. 

\item It has been pointed out by Dine and Anisimov~\cite{DA}, thermal corrections in the the standard two field AD models, (one for inflation and second for AD baryogenesis) can affect the magnitude of  the lepton asymmetry. In those models baryogenesis takes place when the universe is in the Hubble expansion phase with thermal plasma. In contrast, in the case we are considering, we have a single scalar field which does both jobs. 
In these models, when leptogenesis takes place, the universe is not in a thermal phase. Therefore, there are no thermal corrections to the lepton asymmetry

\item We also note that when the phase of the AD field (or the separation between the initial values of the real and the imaginary parts) is large, 
the iso-curvature fluctuations are sufficiently small \cite{Barrie:2021mwi}.

\item We discuss whether this model can be embedded into a higher scale supersymmetric theory since it might appear that the fields $\Phi$, $D,\bar{D}$ and $\chi$ have   resemblance to the superpartner of the right handed neutrino, Higgs doublet and the $U(1)_Y$ gauge field or an  SM singlet superfiield. 
  In a supersymmetrized version of our model, the scalar field $\Phi$  the ``alleged" scalar partner of the right handed neutrino 
  will have to be  lighter than the fermionic component of the superrfield,  the $\nu_R$. 
  However in a typical supersymmetric model,  the
fermionic partners remain light while the scalar partners become heavier due to the addition of SUSY breaking terms. 
Also if the fermion i.e.~the right handed neutrino remains light,  it would have to even under $Z'_2$ like the $\ell$. 
This would then lead to  a tree level mass for the neutrinos via the usual seesaw mechanism. 
Additional symmetries would have to be imposed for preventing this. 
 Also, we have only one $\Phi$ field whereas in a truly SUSY version, there would be three superpartners of 
the three right handed neutrinos.
See for example, a model in  Ref.~\cite{ma} that is a supersymmetric version of the scotogenic model for neutrino mass. This model has some resemblance to ours, although our model is quite different from it in structure and particle content. 
Ref.~\cite{ma} has two $Z_2$ symmetries aside from the continuous lepton number symmetry for preventing the tree level type I seesaw.

\item We also wish to note that in our model the neutrino mass is directly proportional to the $\epsilon$ parameter where the lepton asymmetry is inversely proportional to it. This is very different from the usual leptogenesis mechanism~\cite{FY}, 
where  both neutrino mass and lepton asymmetry are directly proportional to each other.

\item On the phenomenology side, we note that since the lightest active neutrino has a very tiny mass and neutrinos have normal hierarchy, 
we expect the $\langle m_{\beta\beta} \rangle$ parameter in neutrinoiess double beta decay to have a lower bound  on 
$\langle m_{\beta\beta} \rangle > 0.08$ meV. 
This level of neutrinoless double beta decay is of course very hard to achieve with currently planned experiments~\cite{expt}, but nonetheless, it is interesting that there is such a lower limit.

\item Finally, the $D_1$ fermion doublet in our model has a mass in the TeV range to generate the right DM relic density and could therefore be searched for at the colliders. 
It can be produced in a $pp$ collider via the Drell-Yan graph mediated by $W$ exchange with two jet and missing energy signal. The missing energy comes from the $D^\pm$ decaying to $D^0_1$.

\end{enumerate}

\section{Summary} 
We have presented an  extension of the standard model that provides a unified explanation of several of its puzzles i.e.~neutrino masses, dark matter compatible with current direct detection constraints, inflation and baryogegenesis via the Affleck-Dine mechanism. The model adds only three other heavy singlet Majorana fermions ($\chi_i$) and three pairs of $L=0$ SM doublet fermions $D$ and $\bar{D}$, supplemented by  a single lepton number carrying a complex scalar boson, called here the AD field that plays an important role in inflation and baryogenesis.  The lightest of the $D$ and $\bar{D}$ pairs play the role of dark matter. The model  parameters are highly constrained by the requirements of right physics. The interesting point about the model is that AD leptogenesis and WIMP dark matter provide an automatic path to small neutrino mass. We also find it interesting that  the amount of baryon asymmetry in the model is intimately connected to the neutrino mass.
We demonstrate the viability of our model with two sets of benchmark parameters shown in Table II. 
Clearly the model is viable in a domain between these parameters.

\section*{Acknowledgement}

The work of R.N.M. is supported by the US National Science Foundation grant no.~PHY-1914631 and  the work of N.O. is supported by the US Department of Energy grant no.~DE-SC0012447.

\end{document}